\documentclass[sn-mathphys-num]{sn-jnl}

\usepackage{graphicx}%
\usepackage{multirow}%
\usepackage{amsmath,amssymb,amsfonts}%
\usepackage{amsthm}%
\usepackage{mathrsfs}%
\usepackage[title]{appendix}%
\usepackage{xcolor}%
\usepackage{textcomp}%
\usepackage{manyfoot}%
\usepackage{booktabs}%
\usepackage{algorithm}%
\usepackage{algorithmicx}%
\usepackage{algpseudocode}%
\usepackage{listings}%

\begin{document}

\title[Hubbard model]{Hubbard model on a triangular lattice at finite temperatures}

\author{\fnm{Alexei} \sur{Sherman}}\email{alekseis@ut.ee}

\affil{\orgdiv{Institute of Physics}, \orgname{University of Tartu}, \orgaddress{\street{W. Ostwaldi Str 1}, \city{Tartu}, \postcode{50411}, \country{Estonia}}}

\abstract{Using the strong coupling diagram technique, we find three phases of the half-filled isotropic Hubbard model on a triangular lattice at finite temperatures. The weak-interaction ($U\lesssim5t$) and strong-interaction ($U\gtrsim9t$) phases are similar to those obtained by zero-temperature methods -- the former is a metal without perceptible spin excitations; the latter is a Mott insulator with the 120$^\circ$ short-range spin ordering. Zero-temperature approaches predict a nonmagnetic insulating spin-liquid phase sandwiched between these two regions. In our finite-temperature calculations, the Mott gap in the intermediate phase is filled by the Fermi-level peak, which is a manifestation of the bound states of electrons with pronounced spin excitations. We relate the appearance of these excitations at finite temperatures to the Pomeranchuk effect.}


\maketitle

\section{Introduction}
Crystals of organic charge-transfer salts, natrium cobaltate and other layered triangular systems demonstrate several unusual properties \cite{Powell} related to the interplay of pronounced electron correlations and geometric frustrations. The Hubbard model on a triangular lattice is frequently used for the theoretical description of these crystals. Several methods are applied depending on the considered temperature. For the case of zero temperature, they include exact diagonalization \cite{Koretsune}, the path-integral renormalization group \cite{Morita,Yoshioka}, the variational cluster approximation \cite{Sahebsara}, and the density matrix renormalization group \cite{Szasz,Xu}. For finite temperatures, Monte Carlo simulations \cite{Bulut}, the dynamic mean-field theory (DMFT) \cite{Aryanpour}, the dynamic cluster approximation \cite{Lee}, the dual fermion approach \cite{Lee,Li}, and the cellular dynamic mean-field theory \cite{Menke} are used. In the case of an isotropic hopping constant and half-filling, the zero-temperature methods predict the existence of three different regions in the model phase diagram -- a metal without perceptible spin excitations for small Hubbard repulsions, a Mott insulator with a 120$^\circ$ long-range spin order at strong interactions, and a nonmagnetic insulating phase between them. The finite-temperature methods also describe two former phases with the correction that the long-range ordering gives way to the short-range one due to the Mermin-Wagner theorem \cite{Mermin}. The main difference between zero- and finite-temperature approaches concerns the repulsion range corresponding to the intermediate phase. In finite-temperature approximations, states in this range have no Mott gap and are characterized by pronounced magnetic excitations. This part of the phase diagram is attributed to the metallic region in this connection.

In an attempt to clarify this discrepancy, in this work, we investigate the isotropic half-filled Hubbard model on a triangular lattice using the strong coupling diagram technique (SCDT) \cite{Vladimir,Metzner,Pairault,Sherman18}. The method is intended for the case $U\gg t$ and uses the series expansion of Green's functions in powers of the kinetic energy. The approach has several advantages in comparison with the finite-temperature approximations mentioned above. In contrast to the Monte Carlo simulations, SCDT does not suffer from the sign problem and allows one to consider much larger lattices. Unlike DMFT and methods using small clusters, the approach accounts for full-scale charge and spin fluctuations. In contrast with diagrammatic extensions of DMFT, SCDT does not apply vertices obtained for the infinite-dimensional system to the two-dimensional problem. For some ranges of parameters, spins in this problem are ordered, and the correlation length can be large. For the DMFT extensions, such a situation poses the dilemma: vertices of the ordered or disordered DMFT solution should be used. Such a problem does not arise in SCDT.

In this work, we sum up an infinite series of ladder diagrams describing interactions of electrons with charge and spin excitations. The range of the Hubbard repulsions $4t\leq U\leq 12t$ in the lattices up to 24$\times$24 sites is considered. Main results are obtained for the temperature $T\approx0.13t$. Other data derived in the interval $0.06t\lesssim T\lesssim0.32t$ are used for fitting the parameter, ensuring the fulfillment of the Mermin-Wagner theorem. We calculate the local spectral function (LSF), spin and charge susceptibilities, double occupancy $D$, and squared site spin $\langle{\bf S_l^{\rm 2}}\rangle$. We find that the results obtained in the considered ranges of parameters correspond to three qualitatively different types of states. For $U\lesssim5t$, LSFs are mainly concentrated in the narrow peak at the Fermi level (FL) accompanied by weak Hubbard subbands. The zero-frequency spin susceptibility is small and semi-structured as a function of the wave vector. $D$ and $\langle{\bf S_l^{\rm 2}}\rangle$ are close to values  inherent in itinerant electrons. We classify this region as a weakly correlated metal. For $U\gtrsim9t$, states are characterized by the Mott gap at FL in LSF and strong maxima in the zero-frequency spin susceptibility at the $K$ points (the wave vector ${\bf k}=(4\pi/3,0)$ and its symmetric equivalents; here and below the intersite distance is set as the unit of length). $D$ is small and $\langle{\bf S_l^{\rm 2}}\rangle$ is close to its fully localized limit $S(S+1)=0.75$, $S=1/2$. Hence, this region contains states of the Mott insulator characterized by the 120$^\circ$ short-range (for finite $T$) ordering of spins. These two regions are similar to those obtained in the above zero-temperature approaches. However, our calculated states in the intermediate region are metallic, contrasting insulating states obtained in these approaches. Moreover, the LSFs have local maxima at the FL. As in the zero-temperature approaches \cite{Yoshioka,Szasz,Xu}, the states are not entirely nonmagnetic -- intensities of maxima in the zero-frequency spin susceptibility at $K$ points are perceptible though smaller than in the Mott-insulator region. As can be noticed, the strengths of the LSF peaks at the FL grow together with the intensity of the $K$-point susceptibility maxima. This fact points out that the LSF maxima are FL peaks, manifesting the bound states of electrons and spin excitations. By nature, these bound states are similar to the spin-polaron states of the $t$-$J$ model \cite{Schmitt,Ramsak,Sherman94}. The FL peak should be distinct from the Abrikosov-Suhl resonance \cite{Georges}. In SCDT, results of the Anderson impurity model are not used; all calculations are carried out within the framework of the Hubbard model. Earlier, the FL peaks were observed in spectra of the square-lattice Hubbard model \cite{Sherman19}.

The mentioned property of the FL peaks -- their dependence on intensities of the $K$-point susceptibility maxima -- suggests a mechanism that transforms a zero-temperature insulating state to a finite-temperature metallic state in the intermediate region. A system with a nonzero chiral order parameter and moderate spin localization at $T=0$ \cite{Yoshioka,Szasz,Xu} has the potential to gain its entropy by resetting this parameter and increasing spin localization with growing temperature, in analogy with the Pomeranchuk effect in $^3$He \cite{Pomeranchuk,DMLee}. The localized spins form the 120$^\circ$ short-range order with respective spin excitations. They generate bound states with electrons, the manifestation of which -- the FL peak -- fills the Mott gap.

The article is organized as follows: The model Hamiltonian, a brief discussion of the SCDT, and the main formulas are given in the next section. The results of calculations and their discussion are brought up in Sect.~3. The last section is devoted to concluding remarks.

\section{Model and SCDT}
The Hamiltonian of the isotropic Hubbard model on a triangular lattice reads
\begin{equation}\label{Hamiltonian}
H=\sum_{\bf ll'\sigma}t_{\bf ll'}a^\dagger_{\bf l'\sigma}a_{\bf l\sigma}+U\sum_{\bf l}n_{\bf l\uparrow}n_{\bf l\downarrow}-\mu\sum_{\bf l\sigma}n_{\bf l\sigma},
\end{equation}
where {\bf l} and ${\bf l'}$ label sites of a triangular lattice with the base vectors ${\bf a}=(1,0)$ and ${\bf b}=(1/2,\sqrt{3}/2)$, $\sigma=\uparrow,\,\downarrow$ is the spin projection, $t_{\bf ll'}=-t\sum_{\bf c}\delta_{\bf l',l+c}$ with ${\bf c}=\pm{\bf a}$, $\pm{\bf b}$, and $\pm({\bf a-b})$, $a^\dagger_{\bf l\sigma}$ and $a_{\bf l\sigma}$ are electron creation and annihilation operators, $n_{\bf l\sigma}=a^\dagger_{\bf l\sigma}a_{\bf l\sigma}$ is the site occupation operator, and $\mu$ is the chemical potential. In this work, we consider the case of half-filling, $\langle\sum_\sigma n_{\bf l\sigma}\rangle=1$, where the angle brackets denote the statistical averaging with Hamiltonian~(\ref{Hamiltonian}).

As mentioned above, in calculating Green's functions, we use the SCDT series expansion \cite{Vladimir,Metzner,Pairault,Sherman18}. Terms of the expansion are products of the hopping constants and on-site cumulants \cite{Kubo} of the electron creation and annihilation operators. We consider terms with cumulants of the first and second orders only. These cumulants read
\begin{eqnarray*}
&&C^{(1)}(\tau',\tau)=\big\langle{\cal T}\bar{a}_{{\bf l}\sigma}(\tau')a_{{\bf l}\sigma}(\tau)\big\rangle_0,\\
&&C^{(2)}(\tau_1,\sigma_1;\tau_2,\sigma_2;\tau_3,\sigma_3;\tau_4,\sigma_4)\\
&&\quad=\big\langle{\cal T}\bar{a}_{{\bf l}\sigma_1}(\tau_1)a_{{\bf l}\sigma_2}(\tau_2) \bar{a}_{{\bf l}\sigma_3}(\tau_3)a_{{\bf l}\sigma_4}(\tau_4)\big\rangle_0\\
&&\quad\quad-\big\langle{\cal T}\bar{a}_{{\bf l}\sigma_1}(\tau_1)a_{{\bf l}\sigma_2}(\tau_2)\big\rangle_0\big\langle{\cal T}\bar{a}_{{\bf l}\sigma_3}(\tau_3)a_{{\bf l}\sigma_4}(\tau_4)\big\rangle_0\\
&&\quad\quad+\big\langle{\cal T}\bar{a}_{{\bf l}\sigma_1}(\tau_1)a_{{\bf l}\sigma_4}(\tau_4)\big\rangle_0\big\langle{\cal T}\bar{a}_{{\bf l}\sigma_3}(\tau_3)a_{{\bf l}\sigma_2}(\tau_2)\big\rangle_0.
\end{eqnarray*}
The subscript 0 at angle brackets indicates that operator time dependencies and averaging are determined by the site Hamiltonian
\begin{equation*}
H_{\bf l}=\sum_\sigma\big[(U/2)n_{\bf l\sigma}n_{\bf l,-\sigma}-\mu n_{\bf l\sigma}\big].
\end{equation*}
The two last terms of Eq.~(\ref{Hamiltonian}) -- the unperturbed Hamiltonian of the SCDT expansion -- are equal to $\sum_{\bf l}H_{\bf l}$. The symbol ${\cal T}$ is the chronological operator.

The terms of the expansion can be visualized by depicting $t_{\bf ll'}$ as directed lines and cumulants as circles with the number of outgoing and incoming lines equal to the number of creation and destruction operators in them. As in the weak coupling diagram technique \cite{Abrikosov}, the linked-cluster theorem is valid, and partial summations are allowed in SCDT. A two-leg diagram is irreducible if it cannot be divided into two disconnected parts by cutting a hopping line $t_{\bf ll'}$. Denoting the sum of all such diagrams by $K$, the Fourier transform of the electron Green's function $G({\bf l'\tau'},{\bf l\tau})=\langle{\cal T}\bar{a}_{\bf l'\sigma}(\tau')a_{\bf l\sigma}(\tau)$ is written as
\begin{equation}\label{Larkin}
G({\bf k},j)=\big\{\big[K({\bf k},j)\big]^{-1}-t_{\bf k}\big\}^{-1}.
\end{equation}
Here ${\bf k}$ is the 2D wave vector and the integer $j$ defines the fermion Matsubara frequency $\omega_j=(2j-1)\pi T$, $t_{\bf k}$ is the Fourier transform of $t_{\bf ll'}$.

\begin{figure}[t]
\centerline{\resizebox{0.8\columnwidth}{!}{\includegraphics{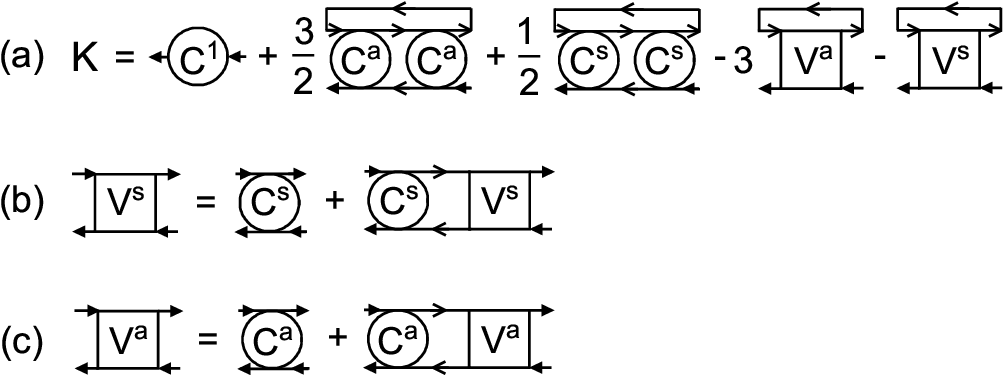}}}
\caption{(a) Diagrams taken into account in calculations of the irreducible part $K({\bf k},j)$. The circle with the notation $C^1$ is the first-order cumulant, circles with notations $C^s$ and $C^a$ are symmetrized and antisymmetrized second-order cumulants, lines with open arrows are renormalized hopping $\theta({\bf k},j)$,  squares with notations $V^s$ and $V^a$ are infinite sums of ladder diagrams symmetrized and antisymmetrized over spin indices. The Bethe-Salpeter equations they satisfy are depicted in parts (b) and (c).} \label{Fig1}
\end{figure}
Diagrams taken into account in the present calculations are shown in Fig.~\ref{Fig1}. Here, block arrows entering and leaving cumulants and vertices shown by squares are their endpoints, lines with open arrows connecting these endpoints are the renormalized hopping,
\begin{equation}\label{theta}
\theta({\bf k},j)=t_{\bf k}+t_{\bf k}^2G({\bf k},j),
\end{equation}
$C^{(s)}$ and $C^{(a)}$ are the second-order cumulants symmetrized and antisymmetrized over their spin indices,
\begin{eqnarray*}
&&C^{(s)}(j+\nu,j,j',j'+\nu)=\sum_{\sigma'}C^{(2)}(j+\nu,\sigma';j,\sigma;j',\sigma; j'+\nu,\sigma'),\\
&&C^{(a)}(j+\nu,j,j',j'+\nu)=\sum_{\sigma'}\sigma\sigma'C^{(2)}(j+\nu,\sigma';j, \sigma; j',\sigma;j'+\nu,\sigma'),
\end{eqnarray*}
with $\nu$ an integer defining the boson Matsubara frequency $\omega_{\nu}=2\nu\pi T$, $V^{(s)}$ and $V^{(a)}$ are analogously symmetrized and antisymmetrized vertices corresponding to sums of infinite sequences of ladder diagrams, which are described by the following Bethe-Salpeter equation (BSE)
\begin{eqnarray}
&&V^{(i)}_{\bf k}(j+\nu,j,j',j'+\nu)=C^{(i)}(j+\nu,j,j',j'+\nu)\nonumber\\
&&\quad+T\sum_{\nu'} C^{(i)}(j+\nu,j+\nu',j'+\nu',j'+\nu){\cal T}_{\bf k}(j+\nu',j'+\nu') \nonumber\\
&&\quad\quad\times V^{(i)}_{\bf k}(j+\nu',j,j',j'+\nu'),\label{BSE}
\end{eqnarray}
where the superscript $i=s$ or $a$. ${\cal T}_{\bf k}(j,j')=N^{-1}\sum_{\bf k'}\theta({\bf k+k'},j)\theta({\bf k'},j')$, and $N$ is the number of sites.

In these notations, the irreducible part $K$ in (\ref{Larkin}) reads
\begin{eqnarray}\label{K}
K({\bf k},j)&=&C^{(1)}(j)+\frac{T^2}{4N}\sum_{{\bf k'}j'\nu}\theta({\bf k'},j') {\cal T}_{\bf k-k'}(j+\nu,j'+\nu)\nonumber\\
&&\times\big[3C^{(a)}(j,j+\nu,j'+\nu,j')C^{(a)}(j+\nu,j,j',j'+\nu)\nonumber\\
&&\quad+C^{(s)}(j,j+\nu,j'+\nu,j')C^{(s)}(j+\nu,j,j',j'+\nu)\big]\nonumber\\
&&-\frac{T}{2N}\sum_{{\bf k'}j'}\theta({\bf k'},j')\big[3V_{\bf k-k'}^{(a)}(j,j,j',j')
+V_{\bf k-k'}^{(s)}(j,j,j',j')\big].
\end{eqnarray}

Expressions for the second-order cumulants appearing in the above equations are rather complex \cite{Vladimir,Metzner,Pairault,Sherman18}. However, in the case
\begin{equation}\label{conditions}
T\ll\mu,\quad T\ll U-\mu,
\end{equation}
they are significantly simplified. This range of chemical potentials corresponds to the considered case of half-filling and moderate doping. In this range, cumulants read
\begin{eqnarray}\label{cumulants}
&&C^{(1)}(j)=\frac{1}{2}\big[g_1(j)+g_2(j)\big],\nonumber \\
&&C^{(2)}(j+\nu,\sigma;j,\sigma';j',\sigma';j'+\nu,\sigma)=\frac{1}{4(T+\zeta)}\big[\delta_{jj'}\big(1-2 \delta_{\sigma\sigma'}\big)\nonumber\\
&&\quad+\delta_{\nu0}\big(2-\delta_{\sigma\sigma'}\big)\big]a_1(j'+\nu)a_1(j)-\frac{1}{2} \delta_{\sigma,-\sigma'}\big[a_1(j'+\nu)a_2(j,j')\\
&&\quad+a_2(j'+\nu,j+\nu)a_1(j)+a_3(j'+\nu,j+\nu)a_4(j,j')\nonumber\\
&&\quad+a_4(j'+\nu,j+\nu)a_3(j,j')\big],\nonumber
\end{eqnarray}
where
\begin{eqnarray*}
&&g_1(j)=({\rm i}\omega_j+\mu)^{-1},\quad g_2(j)=({\rm i}\omega_j+\mu-U)^{-1}, \\
&&a_1(j)=g_1(j)-g_2(j),\quad a_2(j,j')=g_1(j)g_1(j'),\\
&&a_3(j,j')=g_2(j)-g_1(j'),\quad a_4(j,j')=a_1(j)g_2(j').
\end{eqnarray*}
The above approximate formulas somewhat overestimate the tendency to spin ordering, which leads to the transition to the 120$^\circ$ long-range order at finite temperatures for some parameters, in violation of the Mermin-Wagner theorem \cite{Mermin}. The parameter $\zeta$ introduced in Eq.~(\ref{cumulants}) shifts the transition temperature to zero. In the above formulas, it is retained only in cases when multipliers $T^{-1}\delta_{jj'}$ and $T^{-1}\delta_{\nu0}$ in the respective terms are not eliminated by frequency summations. Below, we discuss the way to fix the $\zeta$ value.

With expressions~(\ref{cumulants}), vertices $V^{(s)}$ and $V^{(a)}$ acquire the form
\begin{eqnarray}
&&V_{\bf k}^{(s)}(j+\nu,j,j',j'+\nu)=\frac{1}{2}f_{\bf k}^{(2)}(j+\nu,j'+\nu) \nonumber\\
&&\quad\times\big\{2C^{(s)}(j+\nu,j,j',j'+\nu)-a_2(j'+\nu,j+\nu)z_1({\bf k},j,j')\nonumber\\
&&\quad-a_1(j'+\nu)z_2({\bf k},j,j')-a_4(j'+\nu,j+\nu)z_3({\bf k},j,j')\nonumber\\
&&\quad-a_3(j'+\nu,j+\nu)z_4({\bf k},j,j')\big\},\label{Vs}\\
&&V_{\bf k}^{(a)}(j+\nu,j,j',j'+\nu)=\frac{1}{2}f_{\bf k}^{(1)}(j+\nu,j'+\nu) \nonumber\\
&&\quad\times\big\{2C^{(a)}(j+\nu,j,j',j'+\nu)+\big[a_2(j'+\nu,j+\nu)\nonumber\\
&&\quad-(T+\zeta)^{-1}\delta_{jj'}a_1(j'+\nu)\big]y_1({\bf k},j,j')+a_1(j'+\nu)y_2({\bf k},j,j') \nonumber\\
&&\quad+a_4(j'+\nu,j+\nu)y_3({\bf k},j,j')+a_3(j'+\nu,j+\nu)y_4({\bf k},j,j')\big\},\label{Va}
\end{eqnarray}
where
\begin{eqnarray*}
&&f^{(1)}_{\bf k}(j,j')=\bigg[1+\frac{1}{4}a_1(j)a_1(j'){\cal T}_{\bf k}(j,j')\bigg],\\
&&f^{(2)}_{\bf k}(j,j')=\bigg[1-\frac{3}{4}a_1(j)a_1(j'){\cal T}_{\bf k}(j,j')\bigg],
\end{eqnarray*}
and quantities $z_i$ and $y_i$, $i=1,\ldots 4$ satisfy two systems of linear equations
\begin{eqnarray}
&&z_i({\bf k},j,j')=d_i({\bf k},j,j')-e_{i2}({\bf k},j-j')z_1({\bf k},j,j')\nonumber \\
&&\quad-e_{i1}({\bf k},j-j')z_2({\bf k},j,j')-e_{i4}({\bf k},j-j')z_3({\bf k},j,j')\nonumber\\
&&\quad-e_{i3}({\bf k},j-j')z_4({\bf k},j,j'),\label{zi}\\
&&y_i({\bf k},j,j')=b_i({\bf k},j,j')+\big[c_{i2}({\bf k},j-j')-(T+\zeta)^{-1}\delta_{jj'}c_{i1}({\bf k},j-j')\big]\nonumber \\
&&\quad\times y_1({\bf k},j,j')+c_{i1}({\bf k},j-j')y_2({\bf k},j,j')+e_{i4}({\bf k},j-j')y_3({\bf k},j,j')\nonumber\\
&&\quad+c_{i3}({\bf k},j-j')y_4({\bf k},j,j').\label{yi}
\end{eqnarray}
with the coefficients
\begin{eqnarray*}
&&e_{ii'}({\bf k},\nu)=\frac{T}{2}\sum_j a_i(\nu+j,j)a_{i'}(j,\nu+j){\cal T}_{\bf k}(\nu+j,j) f^{(2)}_{\bf k}(\nu+j,j),\\
&&c_{ii'}({\bf k},\nu)=\frac{T}{2}\sum_j a_i(\nu+j,j)a_{i'}(j,\nu+j){\cal T}_{\bf k}(\nu+j,j) f^{(1)}_{\bf k}(\nu+j,j),\\
&&d_i({\bf k},j,j')=\frac{3}{4}a_i(j,j')a_1(j)a_1(j'){\cal T}_{\bf k}(j,j')f^{(2)}_{\bf k}(j,j')\\ &&\quad-e_{i1}({\bf k},j-j')a_2(j,j')-e_{i2}({\bf k},j-j')a_1(j)\\
&&\quad-e_{i3}({\bf k},j-j')a_4(j,j')-e_{i4}({\bf k},j-j')a_3(j,j'),\\
&&b_i({\bf k},j,j')=-\frac{1}{4}a_i(j,j')a_1(j)a_1(j'){\cal T}_{\bf k}(j,j')f^{(1)}_{\bf k}(j,j')\\ &&\quad+c_{i1}({\bf k},j-j')\big[a_2(j,j')-(T+\zeta)^{-1}\delta_{jj'}a_1(j)\big]+c_{i2}({\bf k},j-j')a_1(j)\\
&&\quad+c_{i3}({\bf k},j-j')a_4(j,j')+c_{i4}({\bf k},j-j')a_3(j,j').
\end{eqnarray*}
Hence two BSEs (\ref{BSE}) for $V^{(s)}$ and $V^{(a)}$ are reduced to two small systems of linear equations (\ref{zi}) and (\ref{yi}), which can be solved exactly for fixed ${\bf k}$, $j$ and $j'$. Vertices $V^{(s)}$ and $V^{(a)}$ describe charge and spin fluctuations, and terms containing them in $K({\bf k},j)$, Eq.~(\ref{K}), allow for their interactions with electrons. The exact solution of the BSEs means that ranges of interactions between electrons and spin and charge excitations, taken into account in calculations, are limited by a considered crystal size only.

Vanishing determinants of the linear systems (\ref{zi}) and (\ref{yi}), $\Delta^{\rm ch}({\bf k},\nu=0)$ and $\Delta^{\rm sp}({\bf k},\nu=0)$, signal the onset of the phase transition, and the symmetry of the emerging phase is defined by the momentum, at which $\Delta\rightarrow0$. In the present problem, $\Delta^{\rm sp}({\bf k},\nu=0)$ vanishes with decreasing $T$ at $K$ points for $U>5t$, which indicates the establishment of the 120$^\circ$ long-range spin order. To avoid the transition at a finite temperature violating the Mermin-Wagner theorem, we set the parameter $\zeta$ to a finite value such that $\Delta^{\rm sp}({\bf k}=(4\pi/3,0),\nu=0)$ vanishes as the temperature approaches 0. This fitting procedure is described in Ref.~\cite{Sherman19a} in application to the square-lattice Hubbard model. Due to frustration, $\zeta$ in the triangular lattice is much smaller than in the square lattice: it equals to zero for $U\leq5t$, $0.06t$ for $5t<U\leq10t$, and $0.08t$ for $U=12t$.

The above equations form a closed set, allowing one to find the electron Green's function by iteration for given values of $U/t$, $T/t$, and $\mu/t$. As the starting function $K({\bf k},j)$ of the iteration, we used $C^{(1)}(j)$, the first term in Eq.~(\ref{K}), which is the irreducible part of the Hubbard-I approximation \cite{Vladimir}.

\section{Results and discussion}
\begin{figure}[t]
\centerline{\resizebox{0.6\columnwidth}{!}{\includegraphics{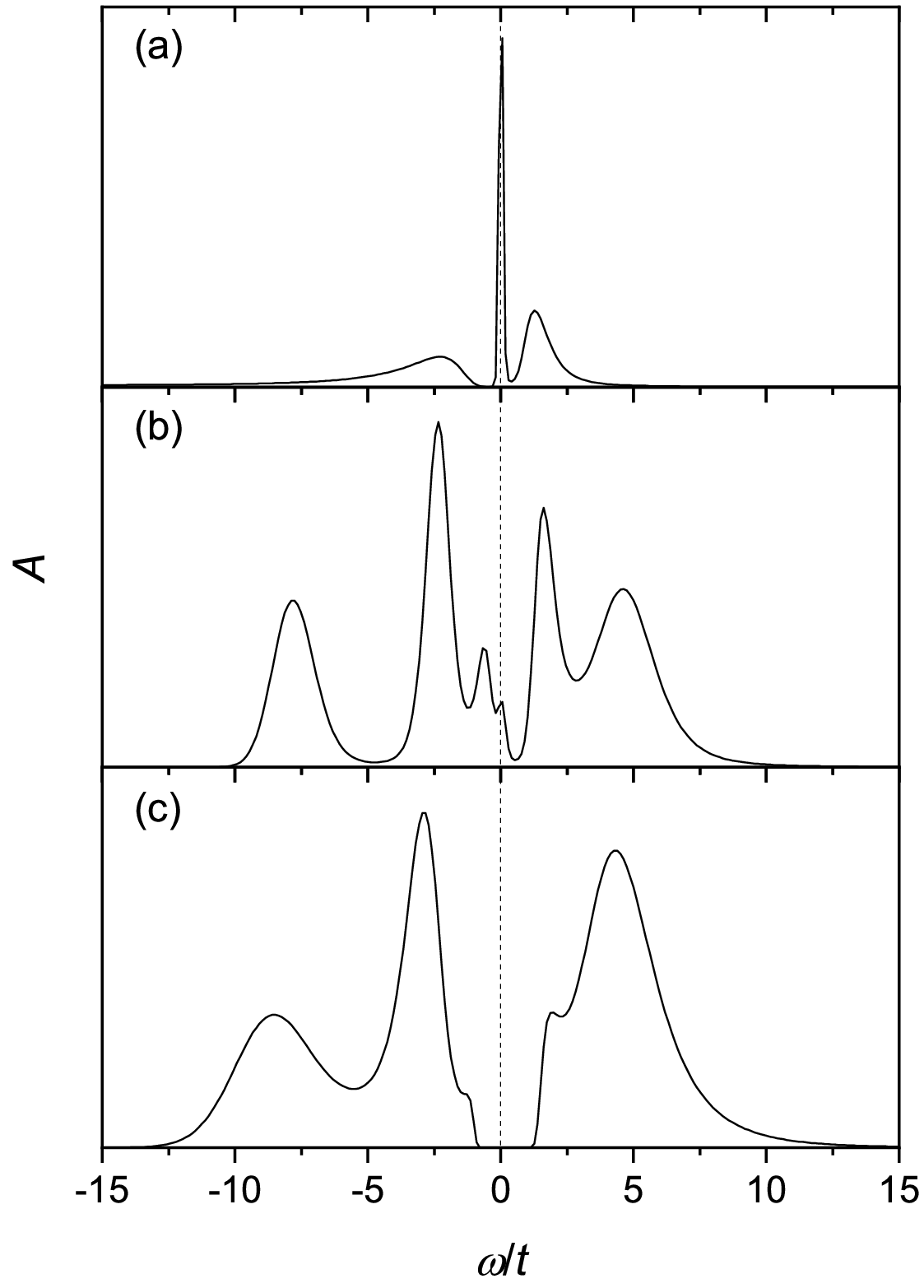}}}
\caption{Calculated local spectral functions for $U=4t$ (a), $8t$ (b), and $10t$ (c) at $T\approx0.13t$.} \label{Fig2}
\end{figure}
Figure~\ref{Fig2} demonstrates calculated electron LSFs,
\begin{equation*}
A(\omega)=-\frac{1}{\pi N}\sum_{\bf k}{\rm Im}\,G({\bf k},\omega),
\end{equation*}
for values of the Hubbard repulsion $U=4t$, $8t$, and $10t$. The analytic continuation to real frequencies $\omega$ was performed using the maximum entropy method \cite{Press,Jarrell,Habershon}. For the smallest $U$, the spectrum has features inherent in a weakly correlated metal -- the strong peak at the FL ($\omega=0$) flanked with weak Hubbard subbands. A similar spectrum is observed for $U=5t$. At $U=8t$, the LSF is characteristic of a strongly correlated metal with pronounced Hubbard subbands. As in the square lattice, there are intensity suppressions near the frequencies $-\mu$ and $U-\mu$ in the subbands (for Fig.~\ref{Fig2}(b), $\mu=4.4t$). They are connected with the multiple electron reabsorption near the Hubbard atom transfer frequencies \cite{Sherman18}. These intensity suppressions lead to the four-band shape of the spectrum, which was much discussed for the square-lattice Hubbard model. Besides these four bands, there are two less intensive features, one on the FL. As seen below, the spectrum would have a gap at the FL without this maximum, and the system would become insulating. Similar LSFs are observed in the repulsion range $5.5t\lesssim U\lesssim 8.5t$. The character of the spectrum is changed for $U\geq9t$ -- a Mott gap opens near FL. An example of such a spectrum is given in Fig.~\ref{Fig2}(c). Hence, in the considered range of $U$, there are three regions with qualitatively different types of LSFs. Spectra in Fig.~\ref{Fig2} were calculated at $T\approx0.13t$. Our data show that three similar areas are also distinguished for lower temperatures, while differences between spectra become less pronounced for higher $T$.

\begin{figure}[t]
\centerline{\resizebox{0.99\columnwidth}{!}{\includegraphics{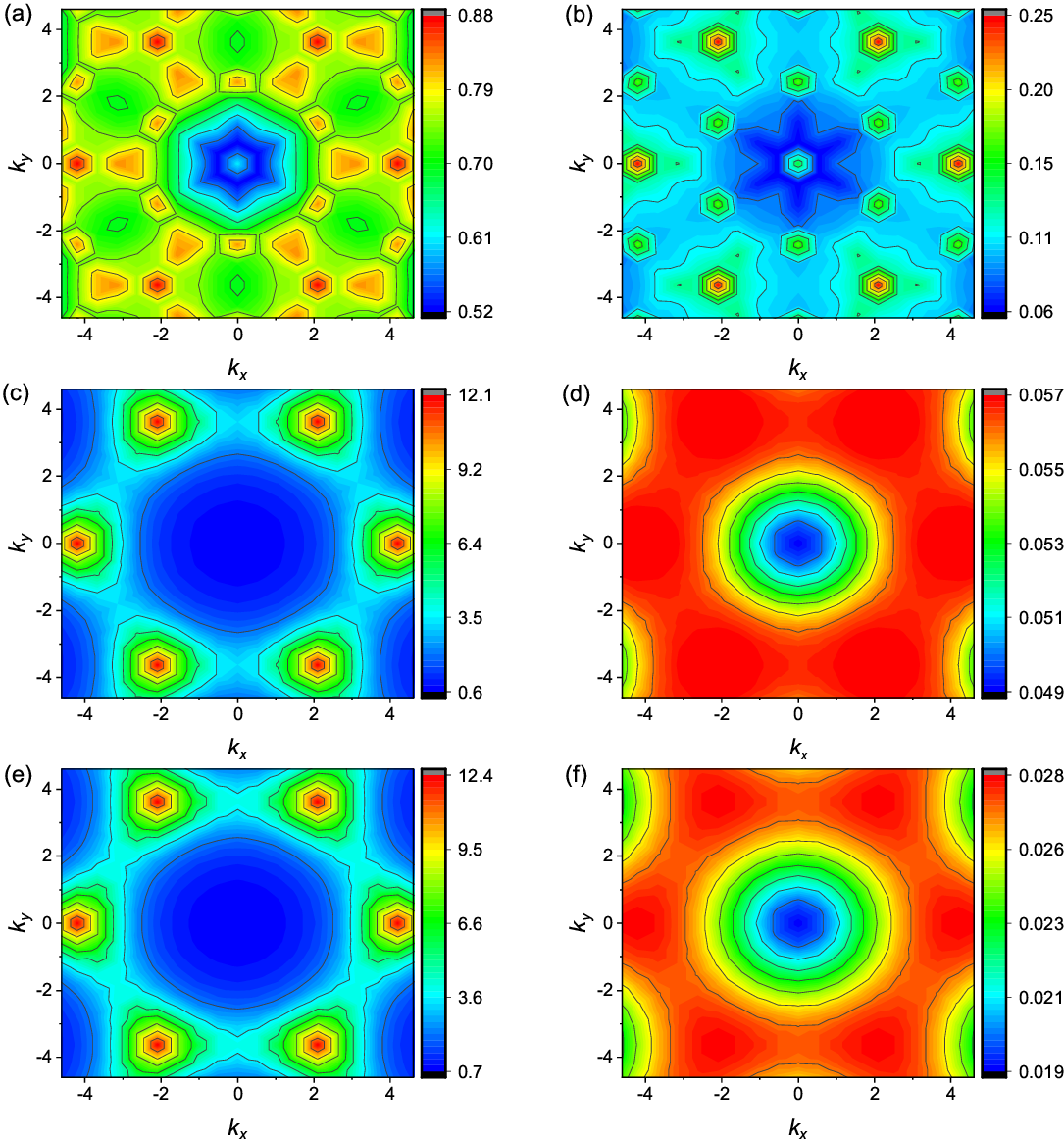}}}
\caption{Contour plots of the zero-frequency spin (left column) and charge (right column) susceptibilities for $U=4t$ (a), (b), $8t$ (c), (d), and $10t$ (e), (f) at $T\approx0.13t$.} \label{Fig3}
\end{figure}

Spin and charge susceptibilities,
\begin{eqnarray*}
&&\chi^{\rm sp}({\bf l'}\tau',{\bf l}\tau)=\langle{\cal T}\bar{a}_{\bf l'\sigma}(\tau')a_{\bf l',-\sigma}(\tau')\bar{a}_{\bf l,-\sigma}(\tau)a_{\bf l\sigma}(\tau)\rangle,\\
&&\chi^{\rm ch}({\bf l'}\tau',{\bf l}\tau)=\frac{1}{2}\langle{\cal T}(n_{\bf l'}(\tau')-\bar{n})(n_{\bf l}(\tau)-\bar{n})\rangle,
\end{eqnarray*}
$n_{\bf l}=\sum_\sigma n_{\bf l\sigma}$, $\bar{n}=\langle n_{\bf l}\rangle$, are expressed in terms of quantities of the previous section as
\begin{eqnarray}
\chi^{\rm sp}({\bf k},\nu)&=&-\frac{T}{N}\sum_{{\bf k'}j}G({\bf k+k'},\nu+j)G({\bf k'},j)-T^2\sum_{jj'} F_{\bf k}(j,\nu+j)\nonumber\\
&&\times F_{\bf k}(j',\nu+j')V_{\bf k}^{(a)}(j+\nu,j'+\nu,j',j),\nonumber\\[-0.5ex]
&&\label{chi}\\[-0.5ex]
\chi^{\rm ch}({\bf k},\nu)&=&-\frac{T}{N}\sum_{{\bf k'}j}G({\bf k+k'},\nu+j)G({\bf k'},j)-T^2\sum_{jj'} F_{\bf k}(j,\nu+j)\nonumber\\
&&\times F_{\bf k}(j',\nu+j')V_{\bf k}^{(s)}(j+\nu,j'+\nu,j',j),\nonumber
\end{eqnarray}
where $F_{\bf k}(j,j')=N^{-1}\sum_{\bf k'}\Pi({\bf k'},j)\Pi({\bf k+k'},j')$ and $\Pi({\bf k},j)=1+t_{\bf k}G({\bf k},j)$. The momentum dependencies of the zero-frequency susceptibilities calculated for parameters of Fig.~\ref{Fig2} are shown in Fig.~\ref{Fig3}.
Both susceptibilities are small and nearly structureless for $U=4t$, which agrees with our classification of this case as a weakly correlated metal. In the intermediate ($U=8t$) and insulating ($U=10t$) regions, the spin susceptibility features strong maxima at $K$ points, which points to pronounced 120$^\circ$ short-range ordering and spin excitations. Strong correlations also reveal themselves in small charge susceptibilities. Shapes and values of susceptibilities for $U=8t$ and $10t$ are very close. Nevertheless, the former is a metal, while the latter is an insulator. As was pointed out above, this difference is connected with the FL peak existing at $U=8t$.

\begin{figure}[t]
\centerline{\resizebox{0.47\columnwidth}{!}{\includegraphics{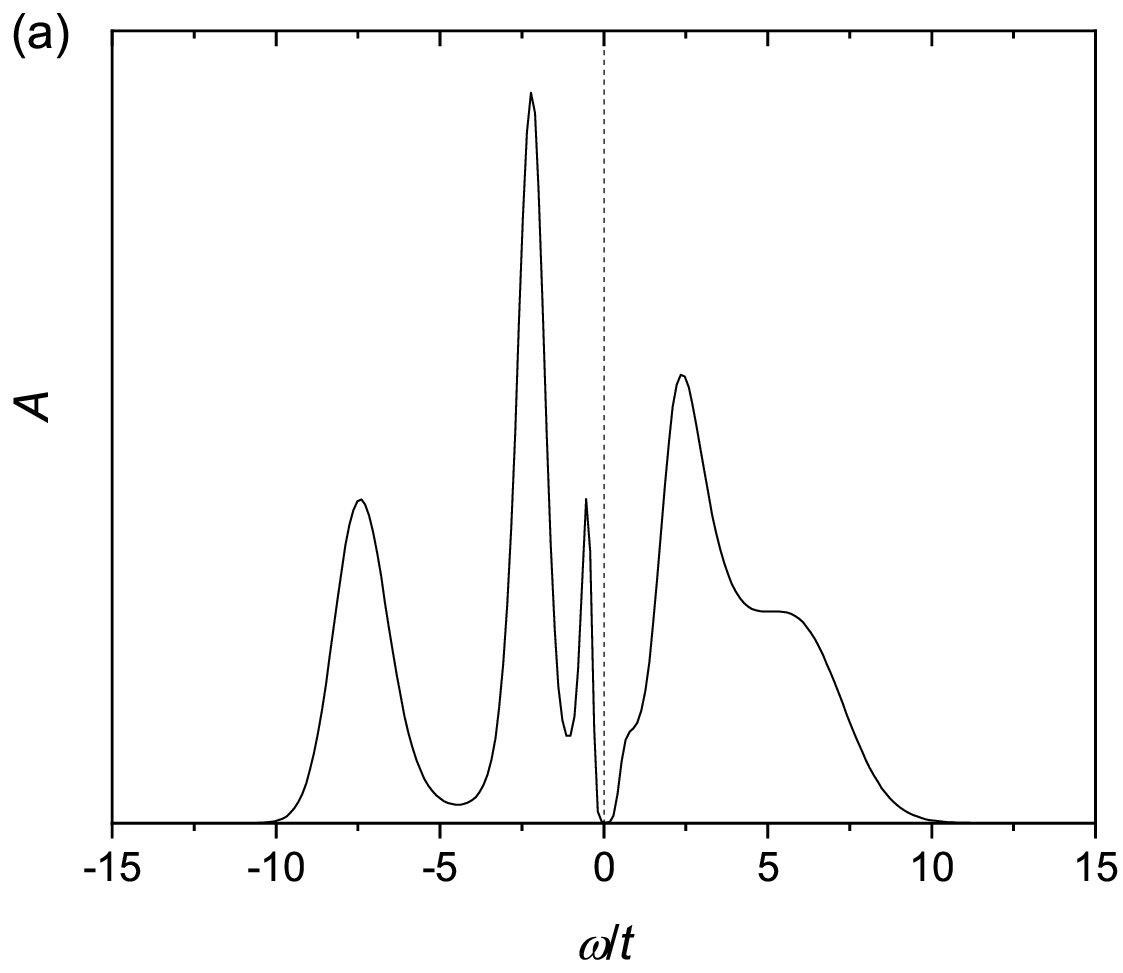}}\hspace{2ex} \resizebox{0.47\columnwidth}{!}{\includegraphics{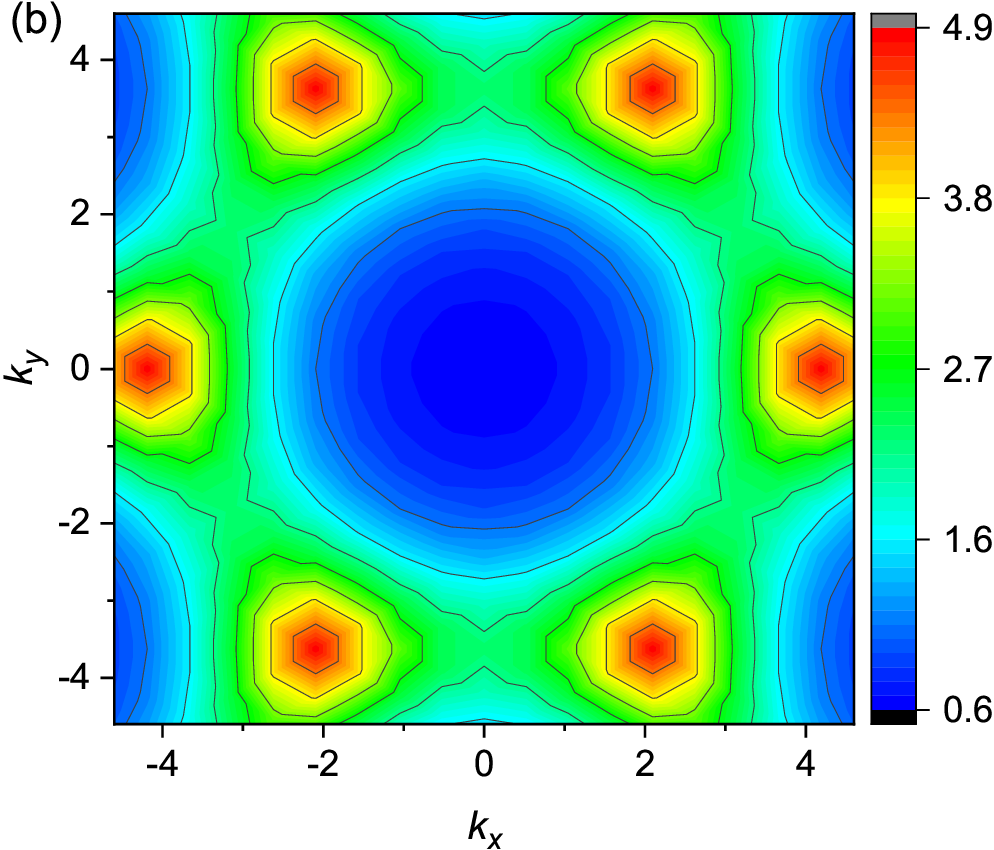}}}
\caption{(a) The local spectral function for $U=8t$, $T\approx0.13t$, and $\zeta=0.12t$. (b) The contour plot of the zero-frequency spin susceptibility for these parameters.} \label{Fig4}
\end{figure}
As mentioned above, equations of the previous section contain parameter $\zeta$, which is fitted such that the transition to the long-range spin order occurs at $T=0$ in agreement with the Mermin-Wagner theorem. For $U=8t$, the fitted value of $\zeta$ is equal $0.06t$. Decreasing or increasing this parameter, we can increase or decrease the strength of maxima at $K$ points in the zero-frequency spin susceptibility. Figure~\ref{Fig4} demonstrates the LSF and zero-frequency spin susceptibility calculated for $U=8t$ and the larger value of the parameter -- $\zeta=0.12t$. Compared with Fig.~\ref{Fig3}(c), intensities of the maxima are reduced more than two times. Consequently, the FL peak disappears, and a Mott gap opens in its place in Fig.~\ref{Fig4}(a). On the one hand, this result justifies that the FL peak is connected with the bound states of electrons and pronounced spin excitations. On the other hand, it shows that the FL peak masks the Mott gap.

The relation between intensities of the susceptibility maxima in Figs.~\ref{Fig3}(e) and \ref{Fig4}(b) is close to that obtained in zero-temperature calculations \cite{Yoshioka,Szasz,Xu} in the ``nonmagnetic'' and spin-ordered insulators. Hence, we could model the result of these calculations to some extent. The susceptibility in Fig.~\ref{Fig3}(c), calculated with fitted $\zeta$ for the intermediate region, has much more pronounced $K$-point maxima in comparison with Fig.~\ref{Fig4}(b) modeling zero-temperature approaches. This result indicates that magnetic momenta are much more localized at finite temperatures than $T=0$. The respective spin excitations are also more pronounced. The mechanism leading to such temperature variation of magnetic moments is known from the physics of $^3$He and the square-lattice Hubbard model \cite{Werner,Fratino,Sherman23}. It is the Pomeranchuk effect \cite{Pomeranchuk,DMLee}, which is connected with the fact that the magnetic moment localization can gain entropy when the temperature exceeds the ordering temperature of the moments. Hence, it is believed that the discrepancy -- insulating character of intermediate states in zero-temperature approaches and their metallic LSF at finite temperatures -- stems from this effect.

\begin{figure}[t]
\centerline{\resizebox{0.98\columnwidth}{!}{\includegraphics{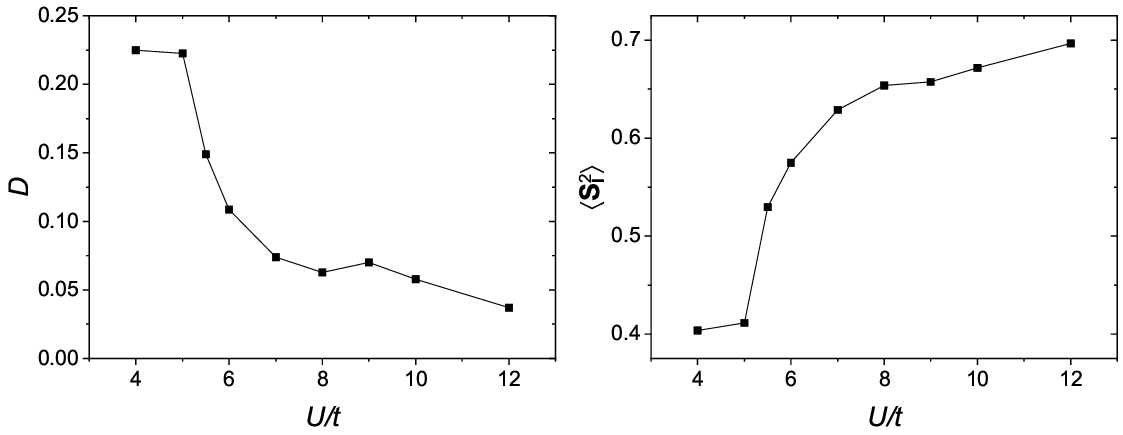}}}
\caption{Dependencies of the double occupancy $D$ (a) and squared site spin $\langle{\bf S}_{\bf l}^2\rangle$ (b) on $U$ for $T\approx0.13t$.} \label{Fig5}
\end{figure}

The double occupancy
\begin{equation*}
D=\langle n_{\bf l\uparrow}n_{\bf l\downarrow}\rangle=\frac{\bar{n}}{2}-\frac{T}{N}\sum_{\bf k\nu}\chi^{\rm sp}({\bf k},\nu)
\end{equation*}
and the squared site spin $\langle{\bf S}_{\bf l}^2\rangle=3\bar{n}/4-3D/2$ are shown in Fig.~\ref{Fig5} as functions of $U$. In agreement with our previous discussion, for $U=4t$ and $5t$ -- the case of a weakly correlated metal -- the double occupancy is close to its itinerant limit of 0.25, and the squared spin is small. Between $U=5t$ and $6t$, the rapid drop of $D$ and the respective growth of $\langle{\bf S}_{\bf l}^2\rangle$ is observed, indicating the transition to the intermediate regime with localized magnetic momenta. The boundary between this and the insulating region reveals itself as a kink on both curves, underlining similarity in properties of the spin subsystem in these two domains. The double occupancy is small for $U=10t$ and $12t$, and the squared local spin is close to its total localization limit $S(S+1)=3/4$ for $S=1/2$.

\section{Concluding remarks}
In this work, we suggested a possible explanation of the qualitative difference in properties of the intermediate phase of the isotropic half-filled Hubbard model on a triangular lattice obtained by zero- and finite-temperature approaches. In the former methods, the phase corresponds to insulating states, while the states are metallic for finite temperature. We carried out calculations using the strong coupling diagram technique. An infinite series of ladder diagrams were taken into account to consider all ranges of interactions of electrons with spin and charge fluctuations in lattices up to 24$\times$24 sites. The interval of Hubbard repulsions $4t\leq U\leq12t$ was studied. Main results were obtained for the temperature $T\approx0.13t$. Other data derived in the range $0.06t\lesssim T\lesssim0.32t$ are used for fitting the parameter $\zeta$, ensuring the fulfillment of the Mermin-Wagner theorem. We calculated local spectral functions, spin and charge susceptibilities, double occupancies, and values of the squared local spin. Obtained states split into three qualitatively different groups. For small repulsions, $U\lesssim5t$, the shapes of calculated LSFs and zero-frequency susceptibilities, small values of $\langle{\bf S}^2_{\bf l}\rangle$, and $D$ close to the itinerant limit 0.25 point to a weakly correlated metal. For large repulsions, $U\gtrsim9t$, the Mott gap at the Fermi level, strong maxima of the zero-frequency spin susceptibility at $K$ points, small values of $D$, and $\langle{\bf S}^2_{\bf l}\rangle$ close to the fully localized spin limit 0.75 indicate a Mott insulator with the short-range spiral order of magnetic momenta. The same properties characterize the intermediate region except for the absence of the Mott gap, which disagrees with the results of zero-temperature approaches. In our finite-temperature calculations, a peak is located on the Fermi level. Similar peaks were observed in the square-lattice Hubbard model for moderate repulsions. They were related to the bound states of electrons with spin excitations -- an analog of the spin-polaron states of the $t$-$J$ model. Indeed, increasing $\zeta$ from the fitted value can decrease the extent of the spin localisation and weaken spin fluctuations. The Fermi-level peak disappears in this case, and a Mott gap opens. On one hand, this confirms the fact that the peak is a manifestation of the bound state of electrons and spin fluctuations. On the other hand, this result points to the fact that the on-site localization of spins is more robust, and, consequently, their fluctuations are more pronounced at finite temperatures than at $T=0$. There is a mechanism that ensures the increase of the magnetic moment localization and the boost of related fluctuations with temperature growth. It is the Pomeranchuk effect -- a system possessing such moments can gain entropy from their on-site localization if the temperature exceeds their ordering temperature. We deal with the two-dimensional system, for which the ordering temperature is zero. Therefore, we suppose that in the intermediate region electron spectra are gaped at zero temperature only. For $T>0$, the bound states of electrons and spin excitations fill the gap.

\backmatter

\section*{Declarations}
Conflict of interest: The author declares no conflict of interest. There are no other applicable declarations.

\end{document}